# TOWARD TEXTUAL INTERNET IMMUNITY

## Gregory M. Dickinson*

*Internet immunity doctrine is broken. Under Section 230 of the Communications Decency Act of 1996, online entities are absolutely immune from lawsuits related to content authored by third parties. The law has been essential to the internet's development over the last twenty years, but it has not kept pace with the times and is now deeply flawed. Democrats demand accountability for online misinformation. Republicans decry politically motivated censorship. And Congress, President Biden, the Department of Justice, and the Federal Communications Commission all have their own plans for reform. Absent from the fray, however—until now—has been the Supreme Court, which has never issued a decision interpreting Section 230. That appears poised to change, however, following Justice Thomas's statement in* Malwarebytes v. Enigma *in which he urges the Court to prune back decades of lower-court precedent to craft a more limited immunity doctrine. This Essay discusses how courts' zealous enforcement of the early internet's free-information ethos gave birth to an expansive immunity doctrine, warns of potential pitfalls to reform, and explores what a narrower, text-focused doctrine might mean for the tech industry.*

## INTRODUCTION

Public concern mounts as a few of the nation's biggest tech players—Google, Amazon, Facebook, and Twitter—dominate the nation's flow of information and online commercial transactions. They facilitate all manner of human activity, for good and for ill, and hold the power to track our movements, guide purchasing decisions, regulate the flow of information, and shape political discourse. Yet, all the while, as private entities, they are free to exercise these powers behind closed doors and, as online rather than physical-world entities, they enjoy immunity from some of the rules that govern their analog counterparts.

With great power has come great controversy. Most recently, Twitter and Facebook have faced criticism for their decisions to disable former President Trump's media accounts[1] and to restrict access to a series of stories published by the *New York Post* about 2020 Democratic presidential candidate Joe Biden's son, Hunter.[2] The decisions are just a few in a stream of high-profile disputes. Last year, Facebook was criticized for its decision *not* to remove a video of Speaker of the House Nancy Pelosi that had been edited to make her appear drunk and confused.[3] And for years now, a debate has been simmering[4] about how the government should respond to various bad-actor websites like those that aid terrorists, facilitate unlawful gun sales,[5] and profit from child abuse and sex trafficking.[6]

---

* Assistant Professor of Law and, by courtesy, Computer Science, St. Thomas University College of Law; Nonresidential Fellow, Stanford Law School Program in Law, Science & Technology; J.D., Harvard Law School.

1. *See* Kevin Roose, *Megaphone to Masses Goes Silent*, N.Y. TIMES, Jan. 8, 2021, at B1 (discussing Twitter's and Facebook's decisions to disable the President's accounts after the attack on the U.S. Capitol); *see also, e.g.*, *Why Ban of @realDonaldTrump Proves Twitter Not 'Just a Platform'*, N.Y. POST (Editorial) (Jan. 8, 2021, 8:00 PM), https://perma.cc/Q6KS-QURQ (arguing that Twitter's decision was politically biased, reasoning that "[s]ome of Trump's tweets were untrue and incendiary, but so are the Ayatollah Khamenei's" and "[h]is account is still up").

## I. SECTION 230 AND ITS CRITICS

At the heart of the controversy lies Section 230 of the Communications Decency Act of 1996,[7] a statute whose tame title belies the weighty protections it provides to the tech industry. Section 230 immunizes online entities against lawsuits related to content created by their users or other third parties: The law promotes "decency" on the internet by allowing online entities to censor "obscene, lewd, lascivious, filthy, excessively violent, harassing, or otherwise objectionable" content without fear of being "treated as the publisher or speaker" of—and held liable—for whatever content they fail to censor;[8] and the law promotes freedom of expression by guaranteeing online entities' ability to relay and host the massive volumes of tweets, snaps, likes, and old-fashioned emails that flow into their systems without incurring liability for their contents.[9] Absent Section 230's protections, online platforms would face an economically crippling duty to review the inconceivable volume of data that flows through their systems to ensure that none of their users' posts contained defamatory speech or other unlawful content.[10] Online platforms might be compelled to heavily censor user speech or disallow online posting altogether to avoid the risk of liability.

---

2. *See* Kevin Roose, *Facebook and Twitter Dodge a 2016 Repeat, and Ignite a 2020 Firestorm*, N.Y. TIMES (Oct. 16, 2020), https://perma.cc/F8BV-K5ZZ (recounting the incident and observing that such incidents put social media companies "in a precarious spot" because they "are criticized when they allow misinformation to spread" and also "when they try to prevent it").

3. *See* Sue Halpern, *Facebook's False Standards for Not Removing a Fake Nancy Pelosi Video*, NEW YORKER (May 28, 2019), https://perma.cc/83NN-WARH.

4. *See, e.g.*, *Justice Department Issues Recommendations for Section 230 Reform*, U.S. DEP'T JUST. (June 17, 2020), https://perma.cc/JYX9-YMZ2 (proposing statutory amendment to address unlawful online activity); *see also* The Eliminating Abusive and Rampant Neglect of Interactive Technologies Act of 2020 (EARN IT Act), S. 3398, 116th Cong. (2020) (a proposed amendment to Section 230 that would allow civil suits against companies that recklessly distribute child pornography).

5. *See* Daniel v. Armslist, LLC, 926 N.W.2d 710 (Wis. 2019), *cert. denied*, 140 S. Ct. 562 (2019); *see also* Dan Frosch & Zusha Elinson, *Texas Man Pleads Guilty in Gun Sale to Mass Shooter*, WALL ST. J. (Oct. 8, 2020), https://perma.cc/93N2-HUTA (describing rising number of individuals who sell guns online, as a business, to individuals they meet through websites like Armslist.com, but who fail to conduct the background checks that are required of frequent private sellers).

6. *See* Doe v. Backpage.com, LLC, 817 F.3d 12 (1st Cir. 2016) (dismissing action brought by underage sex-trafficking victims against website on which they were advertised for sale). So troubled was Congress by the Backpage.com decision that it enacted the Allow States and Victims to Fight Online Sex Trafficking Act (FOSTA), Pub. L. 115-164, 132 Stat. 1253 (2018), which amended Section 230 to specifically exclude sex-trafficking claims from its protections. 47 U.S.C. 230(e)(5). Whether FOSTA helped those Congress sought to protect is an open question. Criminalization of sex work or enhanced enforcement of existing laws may harm sex workers by forcing them to work in secret, more dangerous, environments. *See* Anna North, *Sex Workers Are in Danger. Warren and Sanders are Backing a Bill that Could Help*, VOX (Dec. 17, 2019, 12:20 PM), https://perma.cc/7RUU-36VT.

7. Pub. L. No. 104-104, 110 Stat. 133 (1996) (codified as amended at 47 U.S.C. § 230).

But Section 230—as interpreted by the courts—has not kept pace with the times and now presides over a very different internet from the one it was designed to govern. A law designed to foster free expression now protects entities even should they choose to silence disfavored viewpoints. And, despite its publication-centric roots,[11] Section 230 now insulates online entities from liability for all manner of lawsuits,[12] including product-defect claims—such as the one brought against Snapchat for the design of the app's speed filter,[13] which resulted in many accidents by teenage drivers—and claims against online marketplaces, like the sex-trafficking conspiracy claim brought against the website Backpage.com, which hosted "escort" ads of underage girls and obstructed law-enforcement efforts against sex traffickers so that it could continue to profit from the ad sales.[14]

---

8. 47 U.S.C. § 230(c).

9. *See* 47 U.S.C. § 230(c)(1).

10. *See* Felix T. Wu, *Collateral Censorship and the Limits of Intermediary Immunity*, 87 NOTRE DAME L. REV. 293 (2011) (discussing the risk of censorship by online platforms given differing incentives between speakers, who wish to be heard, and platforms, who are more concerned to avoid legal liability and do not have the resources to review all flagged content).

11. In many ways, Section 230 immunity doctrine merely federalizes common-law defamation principles, under which content distributors like bookstores or libraries are "not required to examine [their publications] to discover whether they contain anything of a defamatory character." RESTATEMENT (SECOND) OF TORTS § 581 cmt. e (AM. L. INST. 1965). Where Section 230, as interpreted by the courts, differs from the common law is its protection even of distributors who have actual knowledge of content's defamatory character yet continue to distribute it. *See* RESTATEMENT (SECOND) OF TORTS § 581 cmt. f (AM. L. INST. 1965) (explaining that a distributor can be liable if it "knows or has reason to know" that content is defamatory); *see also* JOHN C.P. GOLDBERG & BENJAMIN C. ZIPURSKY, RECOGNIZING WRONGS 319–39 (2020) (analyzing Section 230's interaction with pre-internet defamation law and affirmative duties to intervene).

12. *See* Gregory M. Dickinson, *Rebooting Internet Immunity*, 89 GEO. WASH. L. REV. 347, 372–81 (2021) (discussing instances in which Section 230's publication-focused rules are poorly suited for governing the modern internet).

13. Lemmon v. Snap, Inc., 440 F. Supp. 3d 1103 (C.D. Cal. 2020).

14. Doe v. Backpage.com, LLC, 817 F.3d 12, 20-21 (1st Cir. 2016) *superseded by statute*, Allow States and Victims to Fight Online Sex Trafficking Act (FOSTA), Pub. L. 115-164,

Public anger is growing. Not only, it seems, has Big Tech become too powerful, but it even plays by a different set of rules than everyone else. Calls for Section 230 reform have come from every corner. Democrats criticize online platforms' failure to protect the public, reasoning that, given their dominance, online platforms have a responsibility to identify and limit the spread of falsified political ads, hate speech, materials promoting terrorism, and other harmful material.[15] Republicans, for their part, criticize media platforms for their perceived bias, alleging that platforms' content-censorship practices systematically silence conservative voices.[16] And all have come together to criticize Section 230's protection from civil liability of bad-actor websites that purposefully or knowingly facilitate sex trafficking, child pornography, terrorism, and other unlawful activity.[17]

## II. A JUDICIAL CREATION

Somehow spared from criticism, however, has been the judiciary. Big Tech is vilified. Legislative proposals abound. But almost no one has pointed a finger at the courts and judges who are Section 230's true creators. No one, that is, except for Justice Clarence Thomas, who recently reminded us that things could have been—and may still become—otherwise.

Last term, the Supreme Court again declined a chance to interpret Section

---

132 Stat. 1253 (2018).

15. *See* Cecilia Kang, David McCabe & Jack Nicas, *Biden is Expected to Keep Scrutiny of Tech Front and Center*, N.Y. TIMES (Nov. 10, 2020), https://perma.cc/UT79-FNUY (noting that then-president-elect Biden's "clearest position on internet policy" had been to revoke Section 230, which protects tech companies "from lawsuits for hosting or removing harmful or misleading content").

16. *See, e.g.*, Exec. Order No. 13925, 85 Fed. Reg. 34,079 (June 2, 2020) (criticizing platforms' censorship practices as "harming our national discourse" and ordering consideration of rulemaking by the Federal Communications Commission); Online Freedom and Viewpoint Diversity Act, S. 4534, 116th Cong. (2020) (bill sponsored by Republican Senator Roger Wicker that would encourage politically neutral moderation practices by enumerating types of censorship practices eligible for immunity protection); Ending Support for Internet Censorship Act, S. 1914, 116th Cong. (2019) (bill sponsored by Republican Senator Josh Hawley that would make Section 230 immunity contingent on politically neutral content-removal practices).

17. Both Democrats and Republicans have criticized the application of internet immunity doctrine to nonpublication claims and to entities that facilitate unlawful conduct. *See, e.g.*, H.R. REP. NO. 115-572, pt. 1, at 4 (2018) (House Judiciary Committee report on FOSTA noting that "[i]n civil litigation, bad-actor websites have been able to successfully invoke [Section 230] despite engaging in actions that go far beyond publisher functions"); S. REP. NO. 115-199, at 2 (2018) (explaining that Section 230's "protections have been held by courts to shield from civil liability . . . nefarious actors, such as the website BackPage.com").

230, when it denied a request to review the Ninth Circuit's decision in *Malwarebytes v. Enigma*.[18] Despite numerous opportunities to do so,[19] the Court has never interpreted the statute.[20] But that may soon change. Although he agreed with his colleagues' decision not to hear the case, Justice Thomas took the unusual step of issuing a statement to explain why, "in an appropriate case," the Supreme Court should consider the scope of Section 230 immunity.[21] He lamented that lower courts "have long emphasized nontextual arguments when interpreting §230, leaving questionable precedent in their wake."[22] In particular, he questioned courts' application of Section 230 immunity even to platforms that leave content on their sites that they know to be unlawful; to those that seek out and curate unlawful content for their sites; and to claims outside the publishing context, such as those related to defective products.[23] Sensing a gap between Congress's words and current internet immunity doctrine, Justice Thomas urged the Court in a future case to consider whether "the text of [Section 230] aligns with the current state of immunity enjoyed by Internet platforms."[24]

To students of the law, the story is familiar: A statute is stretched by well-meaning judges trying to craft good policy in hard cases, statutory glosses are added to glosses, and eventually the glosses swallow the text to form a doctrine untethered from the statute that gave it life. Such has been the course of internet immunity doctrine under Section 230, whose evolution over the last twenty years has turned the small, unheralded provision attached to the much more comprehensive Communications Decency Act into what can now be fairly called the lynchpin of modern internet law.[25] Its transformation from foundling to foundationary proceeded in two discreet intellectual moves.

First, courts interpreted Section 230's purpose of promoting free expression to operate independently of its promotion of online decency. An entity can claim immunity under the statute for hosting unlawful content even if, rather than slipping through the cracks, the unlawful content is the result of an entity not engaging in any censorship of objectionable material at all. What is more, an entity can claim immunity even if it possesses *actual knowledge* of unlawful material and still fails to remove it.[26] Given that it does nothing to encourage removal of objectionable content, this view is in tension with Section 230's title, "Protection for 'Good Samaritan' blocking and screening of offensive material" and its enactment as part of the "Communications Decency Act."[27] But the approach is not impossible to reconcile with the text and, seemingly more important to courts, it supports a policy of maximal free expression on the internet.

---

18. *See* Malwarebytes, Inc. v. Enigma Software Grp. USA, LLC, 141 S. Ct. 13 (2020) (Thomas, J., concurring in denial of certiorari).
19. *See, e.g.*, Doe v. Backpage.com, LLC, 817 F.3d 12 (1st Cir. 2016), *cert. denied*, 137 S. Ct. 622 (2017); Force v. Facebook, Inc., 934 F.3d 53 (2d Cir. 2019), *cert. denied*, 140 S. Ct. 2761 (2020); Daniel v. Armslist, LLC, 926 N.W.2d 710 (Wis. 2019), *cert. denied*, 140 S. Ct. 562 (2019); Hassell v. Bird, 420 P.3d 776 (Cal. 2018), *cert. denied sub nom*. Hassell v. Yelp, Inc., 139 S. Ct. 940 (2019).
20. *Malwarebytes*, 141 S. Ct. 13, 13.
21. *Id*. at 14.
22. *Id*.
23. *Id*. at 16-18.
24. *Id*. at 14.
25. *See* JEFF KOSSEFF, THE TWENTY-SIX WORDS THAT CREATED THE INTERNET 66–68 (2019) (noting that few media outlets even mentioned Section 230 in their coverage when the Communications Decency Act was enacted).

Courts around the country were concerned that free expression would suffer unless they granted broad Section 230 immunity, even to entities with actual knowledge of unlawful content.[28] They feared what is known as the heckler's veto problem: If platforms become liable for any content they are made aware of but fail to take down, platforms might decide to automatically take down, without investigation, any content even reported to them as objectionable to avoid the cost of investigating.[29] An internet user's post might be taken down and her freedom to speak her mind undermined by the unverified complaint of an internet

---

26. *Malwarebytes*, 141 S. Ct. 13, 15 (explaining that courts have interpreted Section 230 to confer immunity "even when a company distributes content that it *knows* is illegal"); *see, e.g.*, Doe v. Backpage.com, LLC, 817 F.3d 12, 21-24 (1st Cir. 2016) (dismissing claim despite allegations website acted deliberately because "[w]hatever Backpage's motivations, those motivations do not alter the fact that the complaint premises liability on . . . third-party content"); Barnes v. Yahoo!, Inc., 570 F.3d 1096, 1098-1103 (9th Cir. 2009) (dismissing negligence claim where defendant Yahoo acknowledged nude photos posted of plaintiff without her consent but promised to follow through on its promise to remove them); Doe v. Am. Online, Inc., 783 So. 2d 1010, 1018 (Fla. 2001) (dismissing claim where plaintiff alleged AOL was aware that a particular user of its service was transmitting unlawful photographs and yet declined to intervene).

27. *See* Doe v. GTE Corp., 347 F.3d 655, 660 (7th Cir. 2003) (Easterbrook, J.) (noting that if the prevailing approach is correct, "then *§ 230(c)* as a whole makes [online entities] indifferent to the content of information they host or transmit: whether they do (*subsection (c)(2)*) or do not (*subsection (c)(1)*) take precautions, there is no liability under either state or federal law" (emphasis added)); *see generally* ANTONIN SCALIA & BRYAN A. GARNER, READING LAW 176 (2012) (surplusage "canon prevents . . . an interpretation that renders [a provision] pointless").

28. The leading case is the Fourth Circuit's now-famous decision in *Zeran v. America Online, Inc.*, 129 F.3d 327 (4th Cir. 1997) (Wilkinson, C.J.) (AOL immune from suit for its failure to remove, despite notification by plaintiff, messages that purported to advertise for sale shirts with tasteless slogans related to the Oklahoma City bombing and directed individuals to call plaintiff's phone number).

29. *See generally* Brett G. Johnson, *The Heckler's Veto: Using First Amendment Theory and Jurisprudence to Understand Current Audience Reactions Against Controversial Speech*, 21 COMM. L. & POL'Y 175 (2016) (discussing the concept of the heckler's veto, whereby an individual is able to restrict another's freedom to speak by filing complaints against, shouting down, heckling, threatening, or otherwise harassing the speaker); *see also* Reno v. ACLU, 521 U.S. 844, 880 (1997) (invalidating portions of the Communications Decency Act, because, among other reasons, the requirement not to communicate indecent speech to "'specific person[s]' . . . would confer broad powers of censorship, in the form of a 'heckler's veto,' upon any opponent of indecent speech"); Rory Lancman, *Protecting Speech from Private Abridgement: Introducing the Tort of Suppression*, 25 SW. U. L. REV. 223, 253–55 (1996) (discussing

"heckler." To avoid this problem and thereby further a policy of "freedom of speech in the new and burgeoning Internet medium," early courts granted broad immunity under Section 230 to any claim implicating an entity's "exercise of a publisher's traditional editorial functions—such as deciding whether to publish, withdraw, postpone or alter content" even when the entity is made aware that the content is unlawful.[30]

The second, more worrisome, step in Section 230's transformation is its application to nonpublication claims. Although Section 230 is publication centric—it encourages censorship and it speaks in terms of "publishers or speakers" and "content providers"—publication has never been the internet's exclusive function, and it is even less so now than it was in 1996. The internet operates as a virtual world, complete with all manner of goods and services and every kind of wrongdoing. That includes not only publication-related wrongs, like defamation, but also physical-world wrongs, like designing defective smartphone apps or facilitating sex trafficking or illegal gun sales. Claims involving such wrongdoing—against Snapchat, Backpage, or Armslist, for example—raise categorically different issues than Section 230 or the internet immunity doctrine it inspired are designed to handle.[31] Rather than argue that an online entity should have reviewed and moderated third-party content, such claims are analogous to physical-world product defect or conspiracy cases. They argue that an online entity should have designed its app or website differently, typically to include more safety features, or that it intentionally facilitated and profited from unlawful activity.

Courts considering such claims, however, have sometimes ignored the distinction and doubled down on earlier policy-based reasoning. Even for claims that do not allege a failure to review third-party content—and thus do not implicate the moderation burden and heckler's veto concern—courts often grant immunity to defendants on the ground that to do otherwise would interfere with the entity's control over "traditional editorial functions."[32] It does not matter that

---

the origin of the "heckler's veto" concept).

30. *Zeran*, 129 F.3d at 330, 333 (reasoning that even "liability upon notice reinforces service providers' incentives to restrict speech" because the entity would be required to review and investigate so many complaints as to "create an impossible burden").

31. *See* Lemmon v. Snap, Inc., 440 F. Supp. 3d 1103, 1107 (C.D. Cal. 2020) (negligence claim against Snapchat for alleged defects in design of the speed filter feature of its app, which plaintiff alleged contributed to auto accidents); Daniel v. Armslist, LLC, 926 N.W.2d 710, 716 (Wis. 2019), *cert. denied*, 140 S. Ct. 562 (2019) (negligence, public nuisance, wrongful death, aiding and abetting tortious conduct, and other claims against Armslist, alleging that it intentionally designed its website to facilitate illegal gun sales); Doe v. Backpage.com, LLC, 817 F.3d 12, 20 (1st Cir. 2016) (statutory sex-trafficking conspiracy claim against Backpage based on actions site allegedly took to impede law enforcement efforts, such as email anonymization and stripping metadata from photographs).

32. *Zeran*, 129 F.3d at 330; *see also* Malwarebytes, Inc. v. Enigma Software Grp. USA, LLC, 141 S. Ct. 13, 17-18 (2020) (discussing the breadth of protection over entities' editorial decisions and courts' application of immunity in nonpublication contexts).

applying the doctrine to bar product-defect claims is in tension with Section 230's publication focus and the logic of earlier decisions, which premised protection of editorial functions despite a culpable mental state on the need to avoid the heckler's veto. Internet immunity doctrine has become an independent judicial creation, untethered from and largely unconcerned with the words of the statute that gave it life.

Thus, here we are today: a judicially created internet immunity doctrine, a too-powerful tech industry that plays by a different set of rules, and a Supreme Court openly contemplating upsetting the whole house of cards. Of course, it must be acknowledged, much of the dissonance between immunity doctrine and the internet landscape is attributable to the internet's dramatic evolution over the past two decades. Courts could never have foreseen those changes. But that is exactly the problem. By taking a statute targeted to promote internet publication and the censorship of indecent material and pressing it into service as an internet-freedom cure-all, courts have created an expansive doctrine of immunity that is ill-suited for the modern internet, yet now cemented in precedent across the country.

### III. CHARTING A NEW COURSE

That internet immunity doctrine is a judicial creation, however, has its benefits. Courts can always change course. Because the Supreme Court has not (yet) interpreted Section 230, the question of its scope has been left for independent resolution in sixty-three jurisdictions—the thirteen federal circuit courts of appeal and the high courts of the fifty states.[33] Thus far, the story has been one of largely uniform decisions across jurisdictions, all following broadly worded early precedents as if maximal immunity inevitably flows from the words of the statute. It does not, and Justice Thomas's statement in *Malwarebytes* should embolden other courts to say so.

---

33. Although state courts are bound by the Supreme Court on matters of federal law, they are not required to follow decisions of the federal courts of appeal. *Compare* James v. City of Boise, Idaho, 136 S. Ct. 685, 686 (2016) (per curiam) (reversing Idaho Supreme Court decision that it was not bound to follow the Supreme Court's interpretation of a federal statute and explaining that it is the Supreme "Court's responsibility to say what a federal statute means" and "it is the duty of other courts to respect that understanding" (citation and internal quotation marks omitted)), *with* Owsley v. Peyton, 352 F.2d 804, 805 (5th Cir. 1965) ("Though state courts may for policy reasons follow the decisions of the Court of Appeals whose circuit includes their state, they are not obliged to do so." (citation omitted)); *see also* Lockhart v. Fretwell, 506 U.S. 364, 376 (1993) (Thomas, J., concurring) ("[N]either federal supremacy nor any other principle of federal law requires that a state court's interpretation of federal law give way to a (lower) federal court's interpretation."); *Arizonans for Official English v. Arizona*, 520 U.S. 43, 66 n.21 (1997) (noting that "the *stare decisis* effect of [a federal district] court's ruling was distinctly limited" because it was "not binding on the Arizona state courts." (citation and internal quotation marks omitted)). For an intriguing alternative view, that state courts should be bound by lower federal courts' interpretations of federal law, see Amanda Frost, *Inferiority Complex: Should State Courts Follow Lower Federal Court Precedent on the Meaning of Federal Law?*, 68 VAND. L. REV. 53 (2015).

A text-focused renaissance of Section 230 would shift internet immunity doctrine in two directions. First, it would expose tech companies to liability where they act as more than conduits and can be thought of as somehow "responsible . . . for"[34] the content they host—for example, because they know that a piece of unlawful content is on their platforms but fail to remove it or because they intentionally curate it. Second, it would limit online entities' ability to assert immunity in lawsuits not directly related to publication,[35] such as claims for negligence, product defects, conspiracy, or antitrust violations.

Big Tech will inevitably yowl that the destruction of the internet is upon us. It is not. But there is reason for caution. As they develop the contours of what it takes for an online entity to become "responsible for" third-party content or what constitutes a publication-related claim, courts must avoid reimposing on tech companies the same content-moderation burden that drove Congress to enact Section 230 in the first place. Online platforms simply cannot review all the content they host—or even all the complaints they receive—and any claim that would hold them responsible for doing so should remain a nonstarter. That said, Big Tech's parade of horribles about what would happen if courts get it exactly wrong should not deter them from trying to get it right. With public anger growing, Congress inactive, and all branches of government now openly questioning the scope of internet immunity, now is the time for judges to put their judicial laboratories of democracy to work to tailor a textual solution suited for the modern internet.

---

34. 47 U.S.C. § 230(f)(3). Section 230(c)(1) speaks only of "computer services," which are immune from liability, and "information content providers," which are not immune. But the section's definitional subsection, § 230(f)(3), clarifies that information content providers ineligible for immunity includes not only authors of internet content, but also any entity *responsible for* content creation. For analysis of how common-law concepts of vicarious liability might be employed to construe "responsibility," see Agnieszka McPeak, *Platform Immunity Redefined*, 62 WM. & MARY L. REV. 1557, 1584–1612 (2021) (discussing joint enterprise liability); Gregory M. Dickinson, Note, *An Interpretive Framework for Narrower Immunity Under Section 230 of the Communications Decency Act*, 33 HARV. J.L. & PUB. POL'Y 863, 877–81 (2010) (discussing civil conspiracy liability for concert of action). The Ninth Circuit's influential decision in *Roommates.com* initially appeared poised to spur courts in this direction. *See* Fair Hous. Council v. Roommates.com, LLC, 521 F.3d 1157, 1162 (9th Cir. 2008) (en banc) (Kozinski, C.J.) (emphasizing that an online entity becomes a content creator and thus ineligible for immunity if it is "responsible" for "creating or developing" content even "in part"). Subsequent courts have, however, continued to apply immunity broadly, requiring plaintiffs to show an entity's material contribution by "specifically encourag[ing] development of what is offensive about the content" before an entity can be found to have created or developed content. *See* FTC v. Accusearch, Inc., 570 F.3d 1187, 1198–99 (10th Cir. 2009); *see also* Olivier Sylvain, *Intermediary Design Duties*, 50 CONN. L. REV. 203, 258 (2018) (explaining that "in practice," the material contribution standard "makes legal challenges to intermediaries' designs especially difficult to win").

35. Section 230(c)(1)'s text limits its scope to publication-related claims. It bars any lawsuit that would treat an online entity as "the publisher or speaker" of any information provided by a third party. 47 U.S.C. § 230(c)(1). Some courts have limited immunity to causes of action for which publication is a required element. *See, e.g.*, City of Chicago v. StubHub!, Inc., 624 F.3d 363, 366 (7th Cir. 2010) (Easterbrook, C.J.) (reasoning that Section 230 "limits who may be called the publisher of information that appears online," which "might matter to liability for defamation, obscenity, or copyright infringement" but not "Chicago's amusement tax"). Given courts' broad conception of what editorial functions Section 230 protects, however, this requirement is almost always satisfied. *See* Doe v. Backpage.com, LLC, 817 F.3d 12, 19 (1st Cir. 2016) (noting that "[t]he broad construction accorded to section 230 as a whole has resulted in a capacious conception of what it means to treat a website operator as [a] publisher or speaker" and that Section 230 has accordingly been applied to "a wide variety of causes of action, including housing discrimination, negligence, and securities fraud and cyberstalking" (citations omitted)). *But see* Doe v. Internet Brands, Inc., 824 F.3d 846, 853 (9th Cir. 2016) (allowing failure to warn claim to proceed).